# Entrainment flow of emerging jet in a half-space with the no-slip boundary condition

A.V. Gusarov

*Moscow State University of Technology STANKIN, Vadkovsky per. 3a, 127055 Moscow, Russia*

e-mail: av.goussarov@gmail.com

Current development of micro-scale technologies increases the interest to viscous flows with low and moderate Reynolds numbers. This work theoretically studies the entrainment flow of a viscous jet emerging from a plane wall into a half space. Methods are developed to find physically meaningful similarity solutions of the Navier-Stokes equations with finite values of mass and momentum fluxes and having no characteristic size. Two models are considered: flow dominated by momentum flux and flow dominated by mass flux. A new one-parameter set of momentum-dominated similarity solutions is found. Two linearly independent mass-dominated similarity solutions are found. Algorithms are proposed to evaluate the parameters of the similarity models from the mass and momentum balances. Distributions of flow parameters and stresses at the wall are calculated for the similarity models. They are compared with the corresponding distributions obtained by computational fluid dynamics for a more realistic model with a finite size of the jet origin and competitive influence of the mass and momentum fluxes. Such comparison validates the mass-dominated similarity model at the jet Reynolds number $Re \leq 10$ and the momentum-dominated similarity model at $Re \geq 30$. The obtained results are applied to the problem of evaporation at selective laser melting. It is shown that the theoretically estimated flow velocity corresponds to the experimentally observed one. The theory explains formation of the experimentally observed denuded zone and its widening with decreasing the pressure of the ambient gas.

# 1. Introduction

The permanent tendency of miniaturization for devices and technologies is the reason for increasing the interest to viscous flows with low and moderate Reynolds numbers. Studies of nanocapillary flow [1,2] help designing novel membranes. Imaging the external micro-scaled flow at the exit of the nanocapillary can be used for the velocimetry of the internal flow [1-3]. Localized Marangoni convection is responsible for the transport of surfactant [4] and migration of inclusions in a temperature field [5]. Electrically-driven nanocapillary flows are important for microfluidic technologies and understanding the transport of biomolecules [6,7]. Studying the flow field around a copepod is of ecological significance [8]. Analysis of laminar viscous flow is always important for diffusion flames [9]. Viscous microflows play significant role at laser processing of materials [10-12].

In the most of the above-cited works, jet-like viscous flows are analyzed using the well-known exact similarity solution of the Navier-Stokes equations for a concentrated force applied in the unlimited space first obtained by Landau in 1944 [13] (see also more recent publications [14-15]) and later independently by Squire in 1951 [16]. This solution is referred to as Landau jet or Landau-Squire jet. Note that the similarity equations were formulated and reduced to a second-order linear differential equation by Slezkin in 1934 [17]. The contribution of Landau and Squire was finding a particular solution with a clear physical meaning. Yatseyev wrote the general solution of the similarity problem in terms of hypergeometric functions in 1950 [18] and obtained another particular solution. This particular solution was independently found by Squire in 1952 [19] who explained that it describes an emerging jet in a half-space, the so-called aperture flow. The drawback of this solution is that the tangent velocity component at the boundary plane is not zero. Thus, it can not satisfy the no-slip boundary condition typical for a rigid wall.

In 1985, Paull and Pillow [20] analyzed three one-parameter families of similarity solutions of the Navier-Stokes equations. One of them was the Landau jet and the two others were new. Recent theoretical works [21,22] generalize the Landau-Squire jet and propose new applications. Sverak [23] concluded that Landau's one-parameter axisymmetric solution [13-15] describes all possible smooth similarity solutions of the Navier-Stokes equations in unlimited space. Li, Li, and Yan analyzed and classified a two-parameter family of no-swirl solutions with one singularity [24] and a four-parameter family of no-swirl solutions with two isolated singularities [25]. Note, that the general solution of the similarity axisymmetric no-swirl problem given by Yatseyev [18] has exactly four parameters.

In summary, the similarity problem for the Navier-Stokes equations of the viscous flow due to a concentrated force was reduced to a linear ordinary differential equation by Slezkin [17]. The general solution of this equation was formulated by Yatseyev [18]. Landau [14] and Squire [16] independently obtained and analyzed particular solutions for a jet in unlimited space. The Landau-Squire similarity solution has become a powerful tool for analyzing jet flows. Yatseyev [18] and Squire [19] found particular solutions in a half space. The drawback of these solutions is that they do not satisfy the no-slip boundary condition. This makes questionable their applicability to the aperture problem and similar problems of jet emerging into a half-space demanded by current applications. Further theoretical studies [20-25] of this similarity problem have not resulted in a universal approach to the half-space problem with the no-slip boundary condition. The half-space problem was advanced by a different method. Schneider [26] theoretically studied the problem of the jet emerging into a half space in 1981. He distinguished the inner flow near the jet axis and the outer flow. Indeed, the inner flow is essentially the jet and the outer flow is the entrainment flow induced by the jet in the ambient fluid. He concluded that (1) the outer flow is the viscous one with the Reynolds number of the order of one even at very high jet Reynolds numbers of the inner flow and (2) the outer flow can be a similarity flow only in the limit of infinite jet Reynolds number when the flow field becomes singular at the jet axis. He numerically calculated the similarity flow field satisfying the no-slip conditions at the wall with the mass sink at the jet axis estimated for the strong jet of Schlichting [27]. He also combined the numerical solution for the outer flow with the Schlichting analytical approximation for the inner jet to obtain the flow field at moderate jet Reynolds numbers, which was experimentally validated later [28].

In 1985, Schneider [29] revised the half-space problem and calculated the influence of the outer flow on the inner flow. He found that the momentum flux transferred by the inner jet gradually decreases with the distance from the origin and finally vanishes. Thus, at high distances from the origin the flow does not transfer momentum and can not be characterized by an axial force (momentum flux). This is the principal difference from the Landau flow in the unlimited space. Essentially, the concentrated force applied to the gas at the jet origin is totally balanced by the negative pressure of the gas over the wall [29]. Asymptotic expansions [30,31] indicate that at vanishing force, the leading velocity term becomes inversely proportional to the square of the distance from the origin $R$ and can not be described by Landau-type similarity flow with the velocity proportional to $1/R$.

Monographs of Goldshtik, Shtern, and Yavorskii [32,33] considered the asymptotics of the half-space problem in more details. It was concluded that at high distances $R$ from the origin, the whole momentum flux of the jet is balanced by the momentum flux through the wall and the only important quantity is the mass flux. Therefore, the velocity decreases as $1/R^2$ and the Stokes approximation is applicable. However at high jet Reynolds numbers, there exists a range of $R$ where velocity behaves as $1/R$. In this range, the inner and the outer flow regions can be distinguished. The inner and the outer flow fields can be approached by different similarity solutions. Finally, it should be noted that Borchers and Pileckas [34] rigorously proved the $1/R^2$ asymptotics for the half-space problem in 1992.

Briefly, the asymptotics of the half-space problem and the momentum balance were analyzed. However, a similarity solution was only found in the limit of infinitely high Reynolds number. The present work concerns the half-space problem with the no-slip boundary condition. The objective is to propose a tool for analyzing the jet emerging into a half space at arbitrary Reynolds numbers. Section 2 distinguishes two kinds of similarity solutions: jet dominated by momentum flux and jet dominated by mass flux. For the momentum-dominated flow, a method is developed to select the solutions satisfying the no-slip boundary condition from Yatseyev's general similarity solution of the Navier-Stokes equations. For the mass-dominated flow, the general solution is obtained in the Stokes approximation. Section 3 presents the variety of the found similarity solutions and compares them with the results of computational fluid dynamics (CFD) simulation. Section 4 discusses how the similarity solutions can approximate jet flows emerging into a half-space through a finite-size origin (aperture). The related issues of momentum balance and stress distribution over the boundary are raised. An example of application to a micro-scale flow is considered.

## 2. Methods

The jet in a half-space is generally characterized by momentum **F** and mass $M$ fluxes through its origin which can be, for example, an orifice. The focus of the present study is the outer flow around the jet. Therefore, possible specific thermal and kinetic processes near the jet origin are not considered. The fluid is characterized by the fields of density $\rho$, velocity **u** and momentum flow **Π** satisfying the conservation laws for mass and momentum [15],

$$\rho_t + \nabla \cdot (\rho \mathbf{u}) = 0, \quad (\rho \mathbf{u})_t + \nabla \cdot \mathbf{\Pi} = 0, \tag{1}$$

where subscript $t$ designates partial differentiation by time. The constitutive relation for viscous fluid is

$$\mathbf{\Pi} = p\mathbf{I} + \rho \mathbf{u} \otimes \mathbf{u} - \boldsymbol{\sigma}, \tag{2}$$

where $p$ is the pressure, **I** the identity tensor, and **σ** the viscous stress tensor,

$$\boldsymbol{\sigma} = 2\eta \left[ \mathbf{e} - \frac{1}{3}\mathbf{I}\mathrm{tr}(\mathbf{e}) \right], \tag{3}$$

where $\eta$ is the dynamic viscosity and **e** the strain rate tensor defined as the symmetric part of $\nabla \mathbf{u}$. Equations (1)-(3) constitute the common system of the continuity and the Navier-Stokes equations.

## 2.1. Similarity solutions

Steady-state incompressible no-swirl flows without a characteristic size are considered. The vector **F** is supposed to be normal to the wall. Therefore, the flow is axisymmetric. The natural frame for such flow is the polar spherical coordinates with distance from the origin $R$ and polar angle $\theta$. Dimensional analysis [35] proposes the following expression for the flow field:

$$\mathbf{u} = \frac{\nu}{R} \mathbf{U}\left(\theta, \frac{M}{\rho \nu R}, \frac{F}{\rho \nu^2}\right), \quad (4)$$

where $\nu = \eta/\rho$ is the kinematic viscosity and $\mathbf{U}()$ a dimensionless function of dimensionless variables. If the dominant parameter of the jet is momentum flux **F**, the dimensionless function becomes

$$\mathbf{U}\left(\theta, \frac{F}{\rho \nu^2}\right), \quad (5)$$

and velocity decreases as $1/R$ according to Eq. (4). This is the case of the Landau-Squire flow. It is a similarity field because at any $R = \text{const}$, the angular distribution is the same and given by Eq. (5).

If the dominant parameter is mass flux $M$, the dimensionless function is

$$\mathbf{U}\left(\theta, \frac{M}{\rho \nu R}\right), \quad (6)$$

indicating that variables $R$ and $\theta$ do not separate in Eq. (4). Thus, a similarity solution is not generally possible. Consider the case of high viscosity when the second dimensionless variable in Eq. (6) is small and the dimensionless field, Eq. (6), can be expanded in Taylor series about $M/(\rho \nu R) = 0$ reducing the dimensional field, Eq. (4) to

$$\mathbf{u} = \frac{\nu}{R}\left[\mathbf{U}_0(\theta) + \mathbf{U}_1(\theta)\frac{M}{\rho \nu R} + \mathbf{U}_2(\theta)\left(\frac{M}{\rho \nu R}\right)^2 + \ldots\right], \quad (7)$$

with coefficients $\mathbf{U}_i$, $i = 0,1,2,\ldots$. Note that $\mathbf{U}_0 = 0$ because $\mathbf{u} = 0$ at $M = 0$. Thus, the leading term of this expansion is

$$\mathbf{u} = \frac{M\mathbf{U}(\theta)}{\rho R^2}, \quad (8)$$

where index 1 is omitted. This is another similarity velocity field possible in the limit of high viscosity, which decays with distance as $1/R^2$.

For the considered flows in spherical coordinates,

$$\mathbf{u} = \begin{pmatrix} u_R \\ u_\theta \\ 0 \end{pmatrix}, \quad \mathbf{\Pi} = \begin{pmatrix} \Pi_{RR} & \Pi_{R\theta} & 0 \\ \Pi_{R\theta} & \Pi_{\theta\theta} & 0 \\ 0 & 0 & \Pi_{\varphi\varphi} \end{pmatrix}. \quad (9)$$

Equations (1) become

$$\frac{1}{R^2}\frac{\partial(R^2 u_R)}{\partial R} + \frac{1}{R\sin\theta}\frac{\partial(u_\theta \sin\theta)}{\partial\theta} = 0, \tag{10}$$

$$\left.\begin{array}{l}\dfrac{1}{R^2}\dfrac{\partial(R^2 \Pi_{RR})}{\partial R} + \dfrac{1}{R\sin\theta}\dfrac{\partial(\Pi_{R\theta}\sin\theta)}{\partial\theta} - \dfrac{\Pi_{\theta\theta}+\Pi_{\varphi\varphi}}{R} = 0 \quad, \text{ for } R\text{ - component} \\[2mm] \dfrac{1}{R^2}\dfrac{\partial(R^2 \Pi_{R\theta})}{\partial R} + \dfrac{1}{R\sin\theta}\dfrac{\partial(\Pi_{\theta\theta}\sin\theta)}{\partial\theta} - \dfrac{\Pi_{R\theta}-\Pi_{\varphi\varphi}\cot\theta}{R} = 0 \quad, \text{ for } \theta\text{ - component}\end{array}\right\}. \tag{11}$$

The strain rate tensor components are [36]

$$e_{RR} = \frac{\partial u_R}{\partial R}, \quad e_{\theta\theta} = \frac{1}{R}\frac{\partial u_\theta}{\partial\theta} + \frac{u_R}{R}, \quad e_{\varphi\varphi} = \frac{u_R + u_\theta\cot\theta}{R}, \quad e_{R\theta} = \frac{1}{2}\left[R\frac{\partial}{\partial R}\left(\frac{u_\theta}{R}\right) + \frac{1}{R}\frac{\partial u_R}{\partial\theta}\right]. \tag{12}$$

Note that in incompressible flow $tr(\mathbf{e}) = e_{RR} + e_{\theta\theta} + e_{\varphi\varphi} = \nabla\cdot\mathbf{u}$. Therefore, Eq. (3) reduces to

$$\boldsymbol{\sigma} = 2\eta\mathbf{e}. \tag{13}$$

The components of the momentum equation, Eq. (11), become [15]

$$\frac{1}{\rho}\frac{\partial p}{\partial R} + u_R\frac{\partial u_R}{\partial R} + \frac{u_\theta}{R}\frac{\partial u_R}{\partial\theta} = \frac{u_\theta^2}{R} +$$
$$+ \nu\left\{\frac{1}{R^2}\frac{\partial}{\partial R}\left(R^2\frac{\partial u_R}{\partial R}\right) + \frac{1}{R^2\sin\theta}\frac{\partial}{\partial\theta}\left(\frac{\partial u_R}{\partial\theta}\sin\theta\right) - \frac{2}{R^2\sin\theta}\frac{\partial(u_\theta\sin\theta)}{\partial\theta} - \frac{2u_R}{R^2}\right\}, \tag{14}$$

$$u_R\frac{\partial u_\theta}{\partial R} + \frac{1}{\rho R}\frac{\partial p}{\partial\theta} + \frac{u_\theta}{R}\frac{\partial u_\theta}{\partial\theta} + \frac{u_R u_\theta}{R} =$$
$$= \nu\left\{\frac{1}{R^2}\frac{\partial}{\partial R}\left(R^2\frac{\partial u_\theta}{\partial R}\right) + \frac{1}{R^2\sin\theta}\frac{\partial}{\partial\theta}\left(\frac{\partial u_\theta}{\partial\theta}\sin\theta\right) + \frac{2}{R^2}\frac{\partial u_R}{\partial\theta} - \frac{u_\theta}{R^2\sin^2\theta}\right\} \tag{15}$$

Below, partial differential equations (10), (14), and (15) for the velocity components and the pressure are solved.

Consider inverse-power fields with arbitrary positive exponents $m$ and $n$,

$$u_R = \frac{\varphi(\theta)}{R^m}, \quad u_\theta = \frac{f(\theta)}{R^m}, \quad \frac{p}{\rho} = \frac{g(\theta)}{R^n}, \quad . \tag{16}$$

Substituting Eq. (16) into Eqs. (10), (14), (15) results in the system of ordinary differential equations

$$-\frac{ng}{R^{n+1}} + \frac{f\varphi' - f^2 - m\varphi^2}{R^{2m+1}} = \frac{\nu}{R^{m+2}}\left[\frac{(\varphi'\sin\theta)'}{\sin\theta} + (m-1)(m-2)\varphi\right], \tag{17}$$

$$\frac{g'}{R^{n+1}} + \frac{ff' - (m-1)f\varphi}{R^{2m+1}} = \frac{\nu[m\varphi' + m(m-1)f]}{R^{m+2}}, \tag{18}$$

$$\frac{(f\sin\theta)'}{\sin\theta} = (m-2)\varphi, \tag{19}$$

where prime means differentiation by $\theta$. The velocity field can also be expressed through stream function $\psi$ [16],

$$u_R = \frac{1}{R^2 \sin\theta} \frac{\partial \psi}{\partial \theta}, \quad u_\theta = -\frac{1}{R \sin\theta} \frac{\partial \psi}{\partial R}. \tag{20}$$

Equations (20) can be integrated for the inverse-power fields given by Eq. (16) and related by Eq. (19) to obtain

$$\psi = \begin{cases} -f \sin\theta \ln R + \int \varphi \sin\theta \, d\theta, & \text{at } m = 2 \\ -\dfrac{R^{2-m}}{2-m} f \sin\theta, & \text{elsewhere} \end{cases} \tag{21}$$

Substituting Eq. (16) into Eq. (12) and taking into account Eqs. (13) and (2) gives the components of momentum flux

$$\frac{\Pi_{RR}}{\rho} = \frac{g}{R^n} + \frac{\varphi^2}{R^{2m}} + 2\nu \frac{m\varphi}{R^{m+1}}, \tag{22}$$

$$\frac{\Pi_{\theta\theta}}{\rho} = \frac{g}{R^n} + \frac{f^2}{R^{2m}} - 2\nu \frac{f' + \varphi}{R^{m+1}}, \tag{23}$$

$$\frac{\Pi_{\varphi\varphi}}{\rho} = \frac{g}{R^n} - 2\nu \frac{f \cot\theta + \varphi}{R^{m+1}}, \tag{24}$$

$$\frac{\Pi_{R\theta}}{\rho} = \frac{f\varphi}{R^{2m}} + \nu \frac{(m+1)f - \varphi'}{R^{m+1}}, \tag{25}$$

Momentum equations, Eqs. (17) and (18), can be satisfied only if the powers of $R$ in all their terms are equal. It is possible at $m = 1$ and $n = 2$ only. This case corresponds to the well-known Landau-Squire jet with the velocity field decaying as $1/R$. Note that in ideal fluid where the terms proportional to $\nu$ vanish, Eqs. (17) and (18) can be solved at any $n = 2m$. Similarly, in the Stokes approximation at high viscosity the nonlinear terms in the left-hand sides of Eqs. (17) and (18) vanish and solutions at any $n = m + 1$ become possible. The three above cases where $1/R^m$-velocity similarity solutions are possible are summarized in Fig. 1. In all these cases the angular part $\pi(\theta)$ of momentum flux tensor can be separated resulting

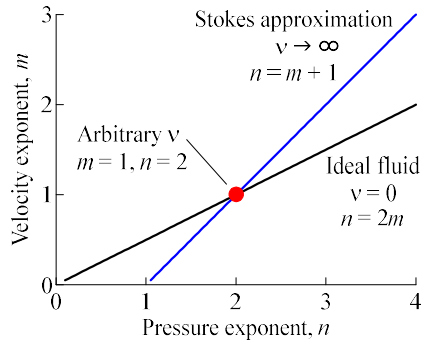

**Fig. 1.** Possible values of velocity $m$ and pressure $n$ exponents in similarity solutions

$$\frac{\Pi}{\rho} = \frac{\pi(\theta)}{R^n}, \tag{26}$$

with the same exponent $n$ as in the expression for pressure in Eqs. (16).

*2.2. Flow dominated by momentum flux (S12 model)*

This is the most important class of similarity solutions with $m = 1$ and $n = 2$ because they satisfy the Navier-Stokes equations at any value of viscosity. It is referred to as S12 model below. Functions $\varphi$ and $f$ introduced by Eqs. (16) are actually the components of vector field $\nu\mathbf{U}$ with dimensionless function $\mathbf{U}$ given by Eq. (5). Note that angular distributions $\varphi$ and $f$ along with $g$ (Eq. (16)) and $\pi$ (Eq. (26)) depend on dimensionless parameter $F/(\rho \nu^2)$ characterizing the intensity of the jet. In this case, the general solution was first obtained by Yatseyev [18]. Equations (27)-(31) below present the sketch of his derivation. Equations (32)-(41) are the general solution itself. Only brief comments are given for Eqs. (27)-(41). The details can be found in the original work [18]. It is a mathematical result and can not be used for applied problems directly. Therefore, the second half of this section proposes a method how to select physically meaningful solutions from Yatseyev's general solution.

System of Eqs. (17)-(19) is reduced to Riccati equation for $f$ [18]

$$f' = \frac{1}{2\nu}f^2 + f\cot\theta + 2\nu\left(\frac{b\cos\theta - a}{\sin^2\theta} + \frac{c}{2}\right), \tag{27}$$

with arbitrary constants $a$, $b$, and $c$. Substitution

$$f = -2\nu\frac{\chi'(\theta)}{\chi(\theta)}, \tag{28}$$

reduces Eqs. (27) to linear equation [18]

$$\chi'' - \chi'\cot\theta + \left(\frac{b\cos\theta - a}{\sin^2\theta} + \frac{c}{2}\right)\chi = 0. \tag{29}$$

Substitution

$$z = \cos^2(\theta/2), \tag{30}$$

transforms Eq. (29) to equation of the Fuchs type [18]

$$\frac{d^2\chi}{dz^2} = \frac{a + b - 2(b+c)z + 2cz^2}{4z^2(1-z)^2}. \tag{31}$$

The general solution of Eq. (31) is [18]

$$\chi(\theta) = c_1\left(\cos\frac{\theta}{2}\right)^\gamma \left(\sin\frac{\theta}{2}\right)^{1+\alpha+\beta-\gamma} F\left(\alpha, \beta; \gamma; \cos^2\frac{\theta}{2}\right) +$$
$$+c_2\left(\cos\frac{\theta}{2}\right)^{2-\gamma} \left(\sin\frac{\theta}{2}\right)^{1+\alpha+\beta-\gamma} F\left(1+\alpha-\gamma, 1+\beta-\gamma; 2-\gamma; \cos^2\frac{\theta}{2}\right), \tag{32}$$

where parameters $\alpha$, $\beta$, and $\gamma$ of hypergeometric function F are related to constants $a$, $b$, and $c$ as

$$\begin{cases} a = \gamma^2 - (1+\alpha+\beta)\gamma + \dfrac{(\alpha+\beta)^2}{2} - \dfrac{1}{2} \\ b = (\alpha+\beta-1)\gamma - \dfrac{(\alpha+\beta)^2}{2} + \dfrac{1}{2} \\ c = \dfrac{(\alpha-\beta)^2 - 1}{2} \end{cases}, \quad (33)$$

Note that the general solution for $\chi$ given by Eqs. (32) and (33) contains five arbitrary constants $a$, $b$, $c$, $c_1$, and $c_2$ while the general solution for $f$ (see Eq. (28)) depends on four parameters only: $a$, $b$, $c$, and $c_2/c_1$ [18].

Function $\varphi$ is directly derived from Eq. (19) as

$$\varphi = -\frac{(f \sin\theta)'}{\sin\theta}. \quad (34)$$

The expression for $g$ was obtained during integration of the system of Eqs. (17)-(19) [18] as

$$g = -2\nu f' + 2\nu^2 \frac{b\cos\theta - a}{\sin^2\theta}. \quad (35)$$

Applying Eq. (27), the derivative of $f$ can be excluded from Eqs. (34) and (35) to obtain

$$\varphi = -\frac{1}{2\nu}f^2 - 2f\cot\theta - 2\nu\left(\frac{b\cos\theta - a}{\sin^2\theta} + \frac{c}{2}\right), \quad (36)$$

$$g = -f^2 - 2\nu f\cot\theta - 2\nu^2\left(\frac{b\cos\theta - a}{\sin^2\theta} + c\right) = \nu\varphi - \frac{1}{2}f^2 - \nu^2 c. \quad (37)$$

The components of the angular part of momentum flow tensor are

$$\pi_{RR} = g + \varphi^2 + 2\nu\varphi = 3g + f^2 + \varphi^2 + 2\nu^2 c, \quad (38)$$

$$\pi_{\theta\theta} = -2\nu^2\left(\frac{b\cos\theta - a}{\sin^2\theta} + c\right), \quad (39)$$

$$\pi_{\varphi\varphi} = 2\nu^2 \frac{b\cos\theta - a}{\sin^2\theta}, \quad (40)$$

$$\pi_{R\theta} = 2\nu^2 \frac{c\cos\theta - b}{\sin\theta}, \quad (41)$$

The general solution given by Eqs. (28), (32), (33), (36)-(41) is applied below to the half-space problem. The outer flow field in a half space is singular at the axis [26,29,32,33]. Velocity components can tend to infinity at $\theta \to 0$ while they are bounded in the interval of polar angle $[\varepsilon, \pi/2]$ at any $0 < \varepsilon < \pi/2$. Important properties of the solution near the axis can be obtained from the conservation laws. Consider mass flux through the hemisphere of radius $R$,

$$\frac{M}{\rho} = 2\pi R^2 \int_0^{\pi/2} u_R \sin\theta \, d\theta = 2\pi R\mu, \quad (42)$$

where integral

$$\mu = \int_0^{\pi/2} \varphi \sin\theta \, d\theta. \tag{43}$$

The mass flux should be the same at any R. According to Eq. (42), it is possible only if $M = 0 = \mu$. Even if integral $\mu$ given by Eq. (43) is improper at $\theta = 0$, it should converge. Suppose that $\varphi$ does not change sign in the vicinity of $\theta = 0$. Then it should satisfy the following condition at the axis:

$$\lim_{\theta \to 0}(\varphi \sin^2 \theta) = 0. \tag{44}$$

Consider axial component of momentum flux through the hemisphere of radius R,

$$F = 2\pi R^2 \int_0^{\pi/2} (\Pi_{RR} \cos\theta - \Pi_{R\theta} \sin\theta) \sin\theta \, d\theta = 2\pi\rho\phi, \tag{45}$$

where integral

$$\phi = \int_0^{\pi/2} (\pi_{RR} \cos\theta - \pi_{R\theta} \sin\theta) \sin\theta \, d\theta = \int_0^{\pi/2} (3g + f^2 + \varphi^2) \cos\theta \sin\theta \, d\theta + 2v^2 b, \tag{46}$$

is evaluated after substituting the expressions for $\pi_{RR}$ and $\pi_{R\theta}$ given by Eqs. (38) and (41). Suppose that $g$ and $f$ do not change sign in the vicinity of $\theta = 0$ too. Then, the sufficient condition for convergence of integral $\phi$ is

$$\lim_{\theta \to 0}(g \sin^2 \theta) = 0 \quad \text{and} \quad \lim_{\theta \to 0}(f \sin\theta) = 0 \quad \text{and} \quad \lim_{\theta \to 0}(\varphi \sin\theta) = 0. \tag{47}$$

The last condition of Eq. (47) for $\varphi$ is stronger than the condition specified by Eq. (44). Below, functions $f$, $\varphi$, and $g$ satisfying the conditions of Eq. (47) at the axis are considered. Multiplying Eqs. (36) or (37) by $\sin^2\theta$ and taking limit as $\theta \to 0$ one can obtain that $a = b$.

No-slip boundary conditions at $\theta = \pi/2$ are

$$f(\pi/2) = 0. \tag{48}$$

$$\varphi(\pi/2) = 0. \tag{49}$$

Consider Eq. (36) at $\theta = \pi/2$,

$$\varphi\left(\frac{\pi}{2}\right) = -\frac{1}{2v} f^2\left(\frac{\pi}{2}\right) + 2v\left(a - \frac{c}{2}\right), \tag{50}$$

It can be consistent with Eqs. (48)-(49) only if $a = c/2$. Finally, three parameters $a$, $b$, and $c$ are related by two necessary conditions for the no-slip flow in a half space

$$a = b = \frac{c}{2}. \tag{51}$$

According to Eq. (50), if Eq. (48) is satisfied along with Eq. (51), Eq. (49) is satisfied too. Thus, Eq. (51) and one of the Eqs. (48) or (49) constitute the sufficient condition for the no-slip flow in a half space. Equation (31) is simplified and its general solution is obtained by substituting Eq. (51) into Eqs. (32)-(33) as

$$\chi(\theta) = c_1 \left(\cos\frac{\theta}{2}\right)^\gamma \left(\sin\frac{\theta}{2}\right)^2 F\left(\alpha, \beta; \gamma; \cos^2\frac{\theta}{2}\right) +$$
$$+ c_2 \left(\cos\frac{\theta}{2}\right)^{2-\gamma} \left(\sin\frac{\theta}{2}\right)^2 F\left(2-\alpha, 2-\beta; 2-\gamma; \cos^2\frac{\theta}{2}\right), \quad (52)$$

with

$$\begin{cases} \alpha = 1 - \frac{1}{2}\sqrt{1+c} + \frac{1}{2}\sqrt{1+2c} \\ \beta = 1 - \frac{1}{2}\sqrt{1+c} - \frac{1}{2}\sqrt{1+2c} \\ \gamma = 1 - \sqrt{1+c} \end{cases} \quad (53)$$

Equations (52) and (53) specify a subset of Yatseyev's general solution [18] given by Eqs. (32)-(33). Equations (28),(36)-(41),(51)-(53) define the solution satisfying the conditions given by Eqs. (47)-(49) if one of the Eqs. (48) or (49) is satisfied, which depends on two parameters: $c$ and $c_2/c_1$.

The similarity solution dominated by momentum flux is determined by the no-slip boundary condition and variable concentrated force $F$. This is why there is no unique solution but a one-parameter set of solutions. Consequently, there should be no unique set of parameters $c$ and $c_2/c_1$ but a functional dependence between them. Note that the solution specified by Eqs. (52) and (53) is not required to satisfy the no-slip boundary condition given by Eqs. (48)-(49). It is Eq. (48) or (49) that define the functional dependence between $c$ and $c_2/c_1$. The left hand side of Eq. (48) $f(\pi/2)$ derived from Eqs. (52) and (53) (see Eq. (28)) is a function of these two parameters. It is explicitly obtained by a computer algebra system. Finally, Eq. (48) is numerically solved by numerical minimizing the absolute value of the left hand side squared, $|f(\pi/2)|^2$. Thus, a solution satisfying Eq. (48) is obtained. Then, Eq. (49) is satisfied automatically (see Eq.(50).

*2.3. Stokes flow dominated by mass flux (S23 model)*

These similarity solutions with $m = 2$ and $n = 3$ describe high-viscosity flows in the Stokes approximation. They correspond to the limit of weak jet dominated by mass flux and referred to as S23 model below. Functions $\varphi$ and $f$ introduced by Eqs. (16) are here the components of vector field $M\mathbf{U}/\rho$ with dimensionless function $\mathbf{U}(\theta)$ (see Eq. (8)). Note that in this approximation, angular distribution of $\varphi, f, g$ (Eq. (16)), and $\pi$ (Eq. (26)) do not depend on mass flux $M$ while the fields of $\mathbf{u}$, $p$, and $\Pi$ are proportional to $M$ (see Eq. (8)). Equations (17)-(19) become

$$\frac{(\varphi'\sin\theta)'}{\sin\theta} + \frac{3g}{\nu} = 0, \quad \frac{g'}{\nu} = 2\varphi' + 2f, \quad (f\sin\theta)' = 0 \quad (54)$$

These equations are integrated with the no-slip boundary conditions, Eqs. (48)-(49).

*2.4. CFD simulation*

The jet flow is modeled by numerical solution of Eqs. (1)-(3) in square $L \times L$ domain of axisymmetric cylindrical coordinates $0 < r < L$, $0 < z < L$ (see Fig. 2). Arbitrary units are used at CFD simulation. They are based on characteristic values of density $\rho_0$, pressure $p_0$, and length $D$. The following parabolic mass flow profile is imposed over the circular jet origin, $z = 0$, $r/D < 1/2$, with the diameter equal to $D$:

$$\rho u_r = 0, \qquad \frac{\rho u_z}{\sqrt{\rho_0 p_0}} = 2\left(1 - 4\frac{r^2}{D^2}\right), \tag{55}$$

with mass flux

$$M = 2\pi \int_0^{1/2} \rho u_z r \mathrm{d}r = \frac{\pi D^2}{4}\sqrt{\rho_0 p_0}. \tag{56}$$

No-slip condition

$$\mathbf{u} = 0, \tag{57}$$

is applied at the rest of the wall, $z = 0$, $r/D > 1/2$. To minimize the influence of external boundaries, $r = L$ and $z = L$, inverse power extensions corresponding to similarity solutions (see Section 2.1),

$$\mathbf{u} \sim \frac{1}{R^m}, \qquad p \sim \frac{1}{R^n}, \tag{58}$$

with $R^2 = r^2 + z^2$, are used as boundary conditions there. Figure 2 summarizes the boundary conditions.

Fluid-dynamic Eqs. (1)-(3) are numerically solved with boundary conditions, Eqs. (55), (57), and (58), by the finite volume method on a uniform $N \times N$ grid implementing a second-order in space and first-order in time Kurganov-Tadmor scheme [37]. Nonlinear artificial compressibility is defined by equation of state

$$\frac{p}{p_0} = \left(\frac{\rho}{\rho_0} - 1\right)\left[1 + 1000\left(\frac{\rho}{\rho_0} - 1\right)^2\right], \tag{59}$$

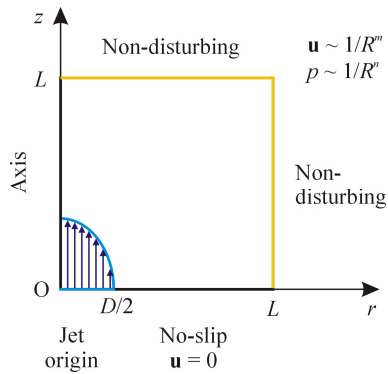

**Fig. 2.** Calculation domain and boundary conditions for CFD simulation

Excluding a small region near the origin with high absolute values of $p$, the deviation of $\rho$ from $\rho_0$ is small. Therefore, the model approximates incompressible flow. A constant value of dynamic viscosity $\eta$ is accepted for each calculation. In the domain of essentially non-compressible flow where $\rho/\rho_0 \approx 1$, kinematic viscosity $\nu \approx \eta/\rho_0$. The Reynolds number is defined as

$$\mathrm{Re} = \frac{\rho_0 <u> D}{\eta} = \frac{D\sqrt{\rho_0 p_0}}{\eta}. \tag{60}$$

where $<u> = 4M/(\pi D^2 \rho_0) = \sqrt{p_0/\rho_0}$ is the mean flow velocity over the jet origin. Steady-state fields **u** and $p$ are obtained by numerical solution. Stream function $\psi$ is defined as

$$\rho u_r = -\frac{1}{r}\frac{\partial \psi}{\partial z}, \qquad \rho u_z = \frac{1}{r}\frac{\partial \psi}{\partial r}. \tag{61}$$

It is calculated by numerical integrating Eqs. (61) with zero initial condition $\psi = 0$ at $r = 0$ and $z = 0$.

## 3. Results

### 3.1. Similarity flow dominated by momentum flux (S12 model)

A one-parameter family of no-slip flows in a half-space is found by the method described in Section 2.2. The similarity solutions are specified by functions of polar angle theta. They can have one or more singularities. If a singularity occurs at $\theta = 0$, it is referred to as the singularity at the axis. A singularity at a non-zero value of $\theta$ is located at a cone around the polar axis. It is referred to as the singularity at a cone. The solution with a singularity at the axis can have finite values of the mass and momentum fluxes. All the considered S12 solutions satisfying the no-slip boundary condition have singularities at the axis. Solutions with a singularity at a cone and finite values of the fluxes are not known. The physical meaning of such solutions is not clear. Therefore they are not studied in this work.

The values of parameters $c$ and $c_2/c_1$ consistent with the no-slip boundary condition are summarized in Appendix A and Fig. 3. For a given value of $c_2/c_1$ ratio, several values of $c$ are generally obtained as shown in Appendix A. Thus, the consistent values of parameters $c$ and $c_2/c_1$ form several continuous branches distinguished in Appendix A by roman numbers. The branches correspond to consecutive spiral coils in the polar plot of $c$ versus $2\mathrm{atan}(c_2/c_1)$ (see Fig.3). The spiral starts at point ($c = -1$, $c_2/c_1 = -1$) shown by an open circle because no solution is found exactly at this point but a solution exists at the spiral at any small distance from this point. The upper bound of $c$ is not found. However at $c > c_0 \approx 15.28938942$, solutions have a singularity at a cone with half-angle $\theta$, $0 < \theta < \pi/2$, and have no finite values of mass and momentum fluxes (see Eqs. (42), (43), (45), and (46)). The physical meaning of such solutions is not clear, and they are not considered below. The spiral has a bit more than two tours and three quarters in interval $-1 < c < c_0$. It intersects radius $c_1/c_2 = 0$ at integer values of $c$ equal to a square of an integer minus one corresponding to integer values of parameter $\gamma$ in Eq. (53). There is no solution at the values of $c = -1, 3, 8$, and 15 (see open circles in Fig. 3). Zero solution $c = f = \varphi = g = 0$ is found at $c_2/c_1 = 0$ (closed circle in Fig. 3) and $c_2/c_1 = -1/2$ (on the spiral). One separate non-zero solution is found at $c \approx -1.93285$ (another closed circle in Fig. 3). Figure 3 proposes that if a solution exists at the given value of $c$, it is unique.

Figure 4 presents new similarity solutions obtained in this work given by Eqs. (28), (36), (37), (51)-(53). The strength of the jet is characterized by the momentum flux $F$ through the hemisphere with the center at the jet origin given by Eqs. (45), (46). Appendix B and Fig. 5 show numerically calculated $F$ as function of constant $c$. The closed point in Fig. 5 designates the solution, which does not belong to the curve (see parameters for this point in Appendix A). The open points on the curve indicate that there is no S12 solution at these points. The indicated points correspond to the points in Fig. 3.

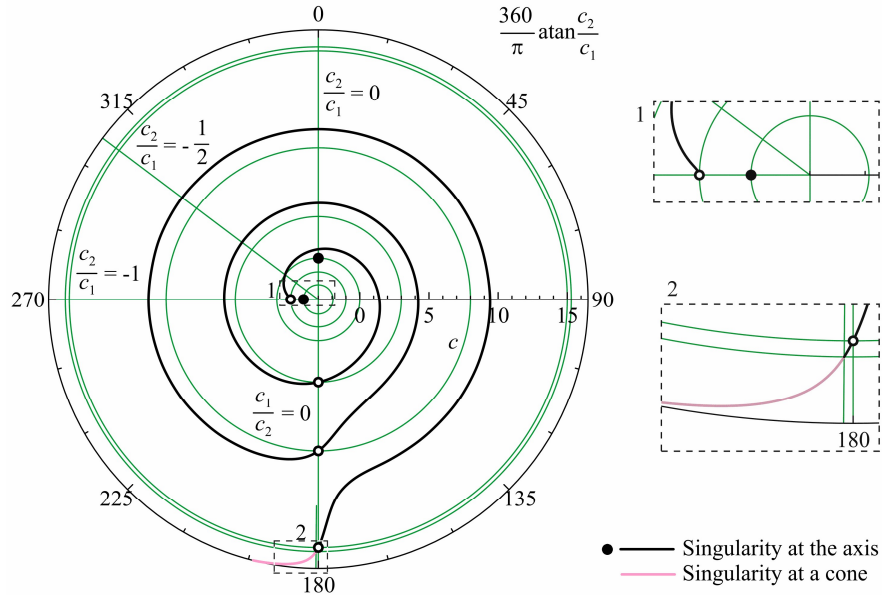

**Fig. 3.** Parameters $c$ and $c_2/c_1$ for momentum-dominated similarity flow fields regular in the half space excluding the axis

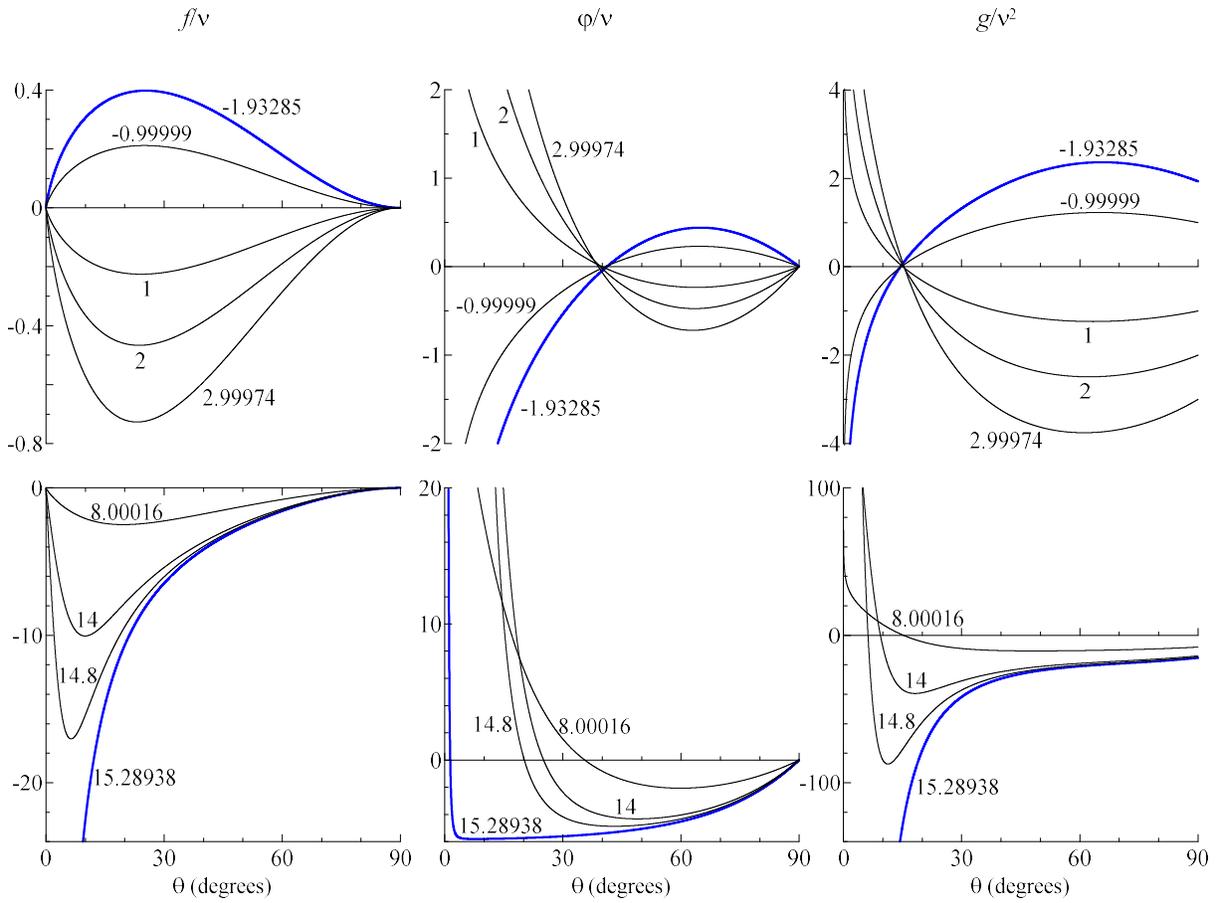

**Fig. 4.** S12 solutions: dimensionless angular profiles of angular $f$ and radial $\varphi$ velocities and pressure $g$ at the values of constant $c$ indicated near the curves

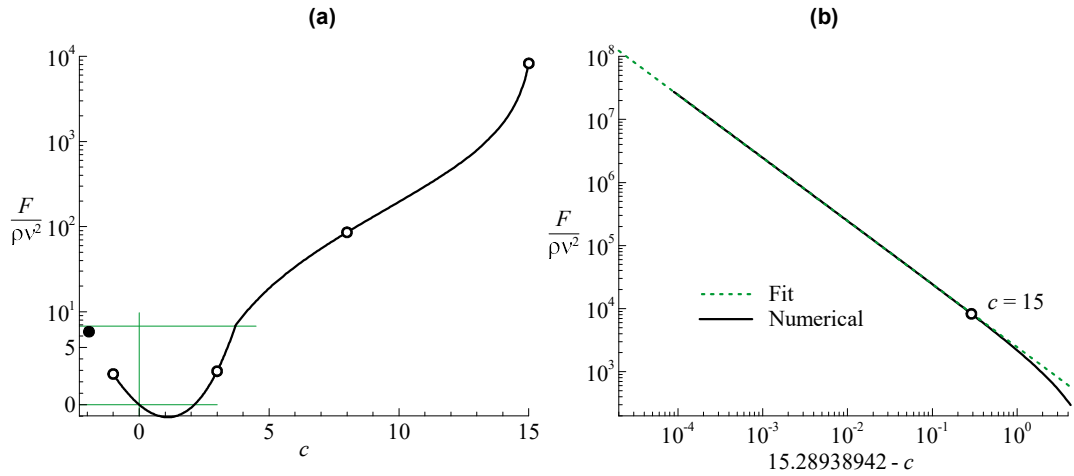

**Fig. 5.** Dimensionless momentum flux $F$ through the hemisphere around the origin for S12 jet flow as function of constant $c$: (a) $c < 15$; (a) $15 < c < c_0$. The closed point designates the solution, which does not belong to the curve. The open points on the curve indicate that there is no solution at these points

## 3.2. Similarity flow dominated by mass flux (S23 model)

The last equation of Eq. (54) is integrated with the no-slip boundary condition, Eq. (48), to obtain

$$f = 0, \qquad (62)$$

indicating that there is no angular velocity component. The components of the angular part of momentum flux (see Eqs. (22)-(25)) become

$$\pi_{RR} = g + 4\nu\varphi, \qquad \pi_{\theta\theta} = \pi_{\varphi\varphi} = g - 2\nu\varphi, \qquad \pi_{R\theta} = -\nu\varphi'. \qquad (63)$$

The integral of the second Eq. (54) indicates that $\pi_{\theta\theta} = \pi_{\varphi\varphi}$ is an arbitrary constant designated below as $\pi_0$,

$$\pi_{\theta\theta} = \pi_{\varphi\varphi} = \pi_0. \qquad (64)$$

Finally, Eq. (64) excludes $g$ from the first Eq. (54) resulting the following equation for $\varphi$:

$$(\varphi'\sin\theta)' + \left(6\varphi + \frac{3\pi_0}{\nu}\right)\sin\theta = 0. \qquad (65)$$

The general solution of Eq. (65) satisfying the no-slip boundary condition, Eq. (49), is

$$\varphi = c_3\left[(3\cos^2\theta - 1)\ln\frac{1+\cos\theta}{1-\cos\theta} - 6\cos\theta\right] - \frac{3\pi_0}{2\nu}\cos^2\theta, \qquad (66)$$

where $c_3$ is another arbitrary constant. The first term of the right hand side of Eq. (66) has a singularity at $\theta = 0$ while the second one is regular. The mass flux through the hemisphere of radius $R$ is

$$\frac{M}{\rho} = 2\pi R^2 \int_0^{\pi/2} u_R \sin\theta \, d\theta = 2\pi\mu, \qquad (67)$$

where integral

$$\mu = \int_0^{\pi/2} \varphi \sin\theta \, d\theta = -\frac{\pi_0}{2\nu} - 2c_3. \qquad (68)$$

## 3.3. CFD simulation

Calculations are made for six different values of viscosity $\eta$ covering the range of Reynolds number (see Eq. (60)) from 2 to 200. Table 1 lists the parameters of the numerical scheme explained in Section 2.4. Exponents $m$ and $n$ are used for the non-disturbing boundary conditions at the outer boundaries of the calculation domain. Resulting recoil force through the jet origin

$$F_o = 2\pi \int_0^{1/2} \Pi_{zz} r \, dr. \qquad (69)$$

is numerically calculated at $z = 0$ to evaluate dimensionless parameter $F/(\rho v^2)$ characterizing jet strength. At Re $\geq$ 30, the numerical solution is compared with similarity model S12 (see Section 2.2). At Re $\leq$ 10, the numerical solution is compared with similarity model S23 (see Section 2.3). The parameters of the similarity solutions are also listed in Table 1. Figure 6 shows the calculated fields for stream function $\psi$, pressure $p$ and radial $u_r$ and axial $u_z$ velocity components. The contours of $\psi$ in Fig. 6 are streamlines.

The purpose of this parametric study is to analyze how the flow varies with the jet Reynolds number Re and to compare S12 and S23 similarity solutions with CFD simulation. Figure 6a shows that the flow qualitatively changes when Re varies in the range from 2 to 30: it is radially divergent at Re = 2, contains circular streamlines at Re = 10, and forms an axial jet and an entrainment flow around it at Re 30. At Re < 30 both streamlines (see Fig. 6a) and pressure contours (see Fig. 6b) approximately correspond to S23 model while at Re ≥ 30 they follow S12 model. At higher Re, the flow is qualitatively the same but the jet becomes narrower with increasing Re (compare the fields at Re = 50, 100, and 200 in Fig. 6). S12 model accurately reproduces the shape of streamlines in the range of Re from 30 to 200 (see Fig. 6a) while a systematic shift of the value of stream function $\psi$ becomes more pronounced with increasing Re because S12 model neglects the mass flux. A considerable deviation of the contours of pressure (see Fig. 6b) and especially axial velocity $u_z$ (Fig. 6d) is observed at Re =100 and 200 in the top part of the shown domain at $z > 10$. Additional CFD computations with various sizes of computation domain $L$ indicated that these deviations are likely due to the influence of the top boundary. Therefore, S12 similarity solution should be more accurate for the half-space problem in these cases.

Table 1. Parameters of CFD calculations and corresponding similarity solutions

| Model | Re | 2 | 10 | 30 | 50 | 100 | 200 |
|---|---|---|---|---|---|---|---|
| CFD | $L/D$ | 32 | 32 | 64 | 16 | 16 | 16 |
| | $N$ | 512 | 1024 | 1024 | 1024 | 1024 | 1024 |
| | $m$ | 2 | 2 | 2 | 1 | 1 | 1 |
| | $n$ | 3 | 3 | 3 | 2 | 2 | 2 |
| | $F_o/(p_0 D^2)$ | 2.54 | 1.06 | 0.98 | 0.98 | 0.98 | 0.98 |
| S12 | $c$ | - | - | 12.53 | 14.26 | 15.04 | 15.23 |
| | $c_2/c_1$ | - | - | 37.27213501 | 118.6442536 | -1813.364896 | -292.3652496 |
| S23 | $-\pi_0(\rho_0/p_0)/D^3$ | 0.125 | 0.05468 | - | - | - | - |
| | $c_3(\rho_0/p_0)^{1/2}/D^2$ | 0 | 0.07419 | - | - | - | - |

**Fig. 6.** (Continued)

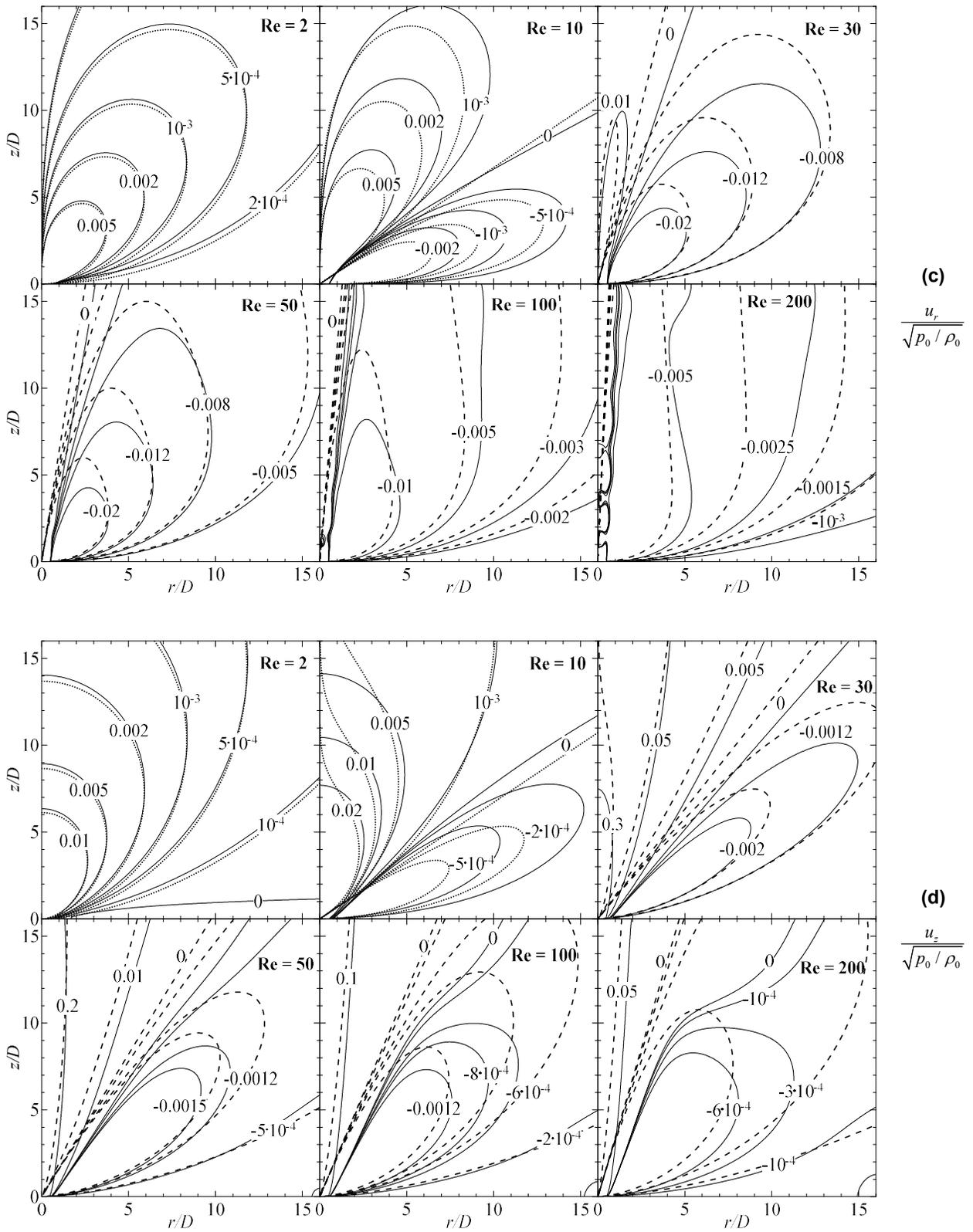

**Fig. 6.** Gas-dynamic fields calculated for CFD (full lines), S12 (dashed lines), and S23 (dotted lines) models: (a) stream function $\psi$, (b) pressure $p$, (c) radial velocity $u_r$, (d) axial velocity $u_z$. The value of Re is indicated at the top right corner of each contour map

## 4. Discussion

*4.1. Validity of CFD simulation*

The numerical CFD results are obtained for the most realistic conditions of finite size of the jet source $D$ and competitive influence of the mass $M$ and momentum $F$ fluxes. Varying the value of $D$ changes the Reynolds number according to Eq. (60). The influence of the Reynolds number on the flow is shown in Fig. 6. The values of the size of computation domain $L$ are taken as high as possible to reduce the influence of the boundaries. The non-disturbing conditions applied at the external boundaries of the computation domain (see Fig. 2 and Eq. (58)) are designed to reduce this influence too. However, it is not always helpful (see the deviation of the contour lines of axial velocity $u_z$ near the top boundary at Re = 100 and 200 in Fig. 6d). According to the obtained experience, no influence of the external boundaries on the CFD results is generally ensured inside the inner square domain ($r = 0 .. L_1, z = 0 .. L_1$) with the size $L_1$ lower than $L$ by a factor of 1.5 .. 2.

The convergence of the numerical scheme was preliminary tested by comparing the fields obtained at different grid sizes $N \times N$. As the result of these tests, the values of $N$ are chosen (see Table 1). The comparison with S12 and S23 similarity solutions is useful outside the domains around the jet origin and those of influence of external boundaries. Good correlation is observed between the CFD results and the corresponding similarity solutions in the mass flux-dominated and momentum flux-dominated modes (see Fig. 6). This also confirms the validity of the CFD computations.

*4.2. Similarity flows*

The general similarity solution for the jet dominated by momentum flux was first formulated by Yatseyev in 1950 [18] (see Eqs.(32) and (33)). However, the physical meaning of the general solution was not analyzed. Up to date, only two particular solutions are known that can be used for applications: the Landau-Squire jet in unlimited space [13,16] and the Yatseyev-Squire jet in a half-space with a slip boundary condition [18,19]. The present work uses Yatseyev's general solution to develop a method for obtaining particular solutions satisfying the no-slip condition in a half space. Section 2.2 describes this method. A new one-parameter set of similarity solutions in a half space is found and described in Section 3.1. Below, the obtained similarity solutions are summarized and the possibility is discussed to apply them for analyzing fluid flows at realistic conditions.

The present work analyses only the solutions satisfying the no-slip boundary condition in a half space and having finite values of the mass and momentum flux. These solutions are given by Eqs. (52) and (53). Appendixes A and B and Figs. 3-5 show the results obtained by numerical calculation starting from these equations. The results indicate that the found set of solutions is one-parametric. Solutions exist at the values of constant $c$ approximately equal to -1.93285 and in the range from -1 to $c_0 \approx 15.28938942$ excluding the values of -1, 3, 8 and 15. The results propose that if a solution exists at the given value of $c$, it is unique. Constant $c$ is directly related with the distributions of normal $\sigma$ and shear $\tau$ stresses over the wall. They are calculated as $\sigma = \Pi_{\theta\theta}$ and $\tau = \Pi_{R\theta}$ at $\theta = \pi/2$:

$$\sigma = \tau = -c\frac{\rho v^2}{R^2}. \tag{70}$$

At positive $c$, the pressure near the wall is negative (see the right column in Fig. 4) and the tangential velocity near the wall (see the middle column in Fig. 4) is directed to the center. This is why both $\sigma$ and $\tau$ are negative. At negative $c$, the flow direction and the sign of pressure change (see Fig. 4), and $\sigma$ and $\tau$ become positive.

All the found non-zero S12 solutions have singularity at the axis: the axial velocity (see the middle column in Fig. 4) and pressure (see the right column in Fig. 4) have no finite values at $\theta = 0$. The axial singularity is commonly interpreted as the strong jet opposed to the weak entrainment flow around it [26,30,32,33]. Moreover, the existing methods to evaluate the flow field theoretically are based on the separate consideration of the slender Schlichting-type jet acting as a linear mass sink for the outer flow [26]. Such approximate solutions converge at increasing the jet strength (Reynolds number). The present work proposes the method to obtain exact solutions of the Navier-Stokes equations in a half-space with no-slip boundary conditions applicable to both strong and weak jets. The solution is valid both for the inner jet flow and the outer entrainment flow. The approximate solution gave the maximum value of $c$, $c_0 = 15.2894$ [32,33]. Within the reported digits, it is the same value as found in the present work from the exact solution.

Figure 4 shows that the flow qualitatively changes with the sign of $c$: it is the jet emerging from a wall at $c > 0$ and the jet impinging on a wall at $c < 0$. The absolute values of velocity and pressure generally increase with the absolute value of $c$. Momentum flux $F$ shown in Fig. 5 seems to contradict the character of flow because $F$ is positive for the impinging jet when $c$ is negative and, on the contrary, $F$ is negative for the emerging jet at positive values of $c$ in the range from 0 to 2 (see Fig. 5). The issue is that $F$ takes into account the momentum flux through the hemisphere (see Eq. (45)) while the normal stress at the wall (see Eq. (70)) contributes to the momentum balance too. In addition, the axial singularity is a linear source of momentum. These additional contributions to the momentum balance are especially important at small $c$. Starting from Eq. (41), one can evaluate the linear force per unit length of the axis as

$$F_l = \lim_{\theta \to 0}(2\pi R \sin\theta \Pi_{R\theta}) = c\frac{2\pi\rho v^2}{R}. \tag{71}$$

Consider the momentum balance in the flow domain limited by two hemispheres with radii $R_1$ and $R_2$ as shown in Fig. 7a. The momentum flux $F$ through the hemisphere is independent of the radius. Therefore, flux $F$ entering into the domain through the inferior hemisphere is balanced by the same flux exiting through the superior hemisphere. The momentum arising at the axis per unit time in the interval from $R_1$ to $R_2$ is equal to the momentum flux exiting through the ring of the wall limited by circles with radii $R_1$ and $R_2$,

$$J = \int_{R_1}^{R_2} F_l dR = -2\pi \int_{R_1}^{R_2} \sigma R dR = 2\pi \rho v^2 c \ln \frac{R_2}{R_1}. \tag{72}$$

Therefore, flux $J$ entering into the domain through the axis is balanced by the same flux exiting through the wall. The value of $J$ is always positive for emerging jets with $c > 0$ and negative for impinging jets with $c < 0$. Figure 7a summarizes the momentum fluxes. According to this figure, the intensity of the jet flow is better characterized by the sum of momentum fluxes $F + J$. If the logarithm in Eq. (72) is of the order of one, $J$ becomes the principal term of the sum at small $c$. Therefore, $F + J$ is positive for emerging jets with $c > 0$ and negative for impinging jets with $c < 0$.

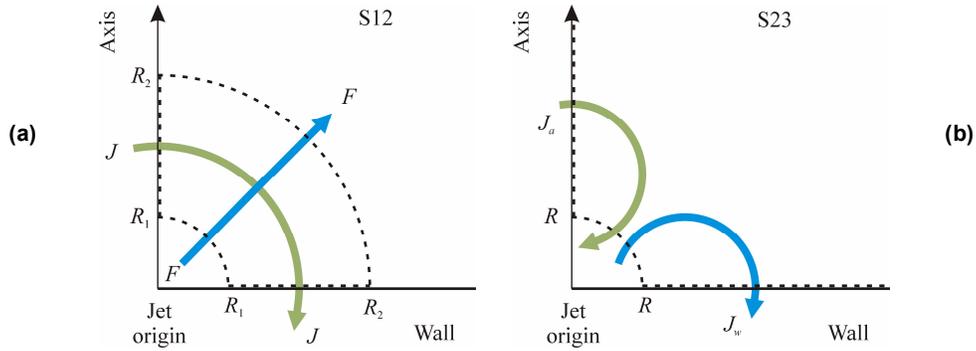

**Fig. 7.** Momentum balance at similarity jet flows in a half space: (a) jet dominated by momentum flux (S12 model); (b) jet dominated by mass flux (S23 model). Arrows indicate the directions of momentum transfer

In the case of jet dominated by mass flux (S23 model), momentum flux through the hemisphere depends on radius $R$,

$$F = 2\pi R^2 \int_0^{\pi/2} (\Pi_{RR} \cos\theta - \Pi_{R\theta} \sin\theta) \sin\theta d\theta = \frac{2\pi \rho \phi}{R}, \tag{73}$$

where integral

$$\phi = \int_0^{\pi/2} (\pi_{RR} \cos\theta - \pi_{R\theta} \sin\theta) \sin\theta d\theta = -\pi_0 - 4vc_3, \tag{74}$$

is calculated by substituting the general solution given by Eqs. (62)-(64) and (66). Linear momentum source at the axis and momentum flux through the wall are also important for the momentum balance. The linear force per unit length of the axis is

$$F_l = \lim_{\theta \to 0}(2\pi R \sin\theta \Pi_{R\theta}) = \frac{8\pi\rho v c_3}{R^2}. \qquad (75)$$

The momentum arising at the axis per unit time at the distances from the origin greater than $R$ is

$$J_a = \int_R^\infty F_l dR = \frac{8\pi\rho v c_3}{R}. \qquad (76)$$

The normal stress at the wall is

$$\sigma = \frac{\rho \pi_0}{R^3}. \qquad (77)$$

The momentum flux from the fluid through the wall domain excluding the disk of radius $R$ is

$$J_w = -2\pi \int_R^\infty \sigma R dR = -\frac{2\pi\rho\pi_0}{R}. \qquad (78)$$

Note that the momentum flux through the hemisphere given by Eqs. (73) and (74) can be decomposed as

$$F = J_w - J_a, \qquad (79)$$

Consider momentum balance in the infinite domain formed by the half space excluding the hemisphere of radius $R$ shown in Fig. 7b. Equation (79) means that flux $J_a$ enters into the domain through the axis and exits through the hemisphere and flux $J_w$ enters through the hemisphere and exits through the wall.

In summary, similarity flows in a half-space with the no-slip boundary condition are studied in detail. The methods proposed in this work can be generalized for conical domains with arbitrary half-angles and for other boundary conditions. Such flows can be the subject of further researches.

*4.3. Approximation of flow fields by similarity solutions*

Jet flow is essentially characterized by mass $M$ and momentum $F_o$ fluxes through the origin (for example, an orifice or an evaporation spot) while the parameters of the considered similarity solutions are constant $c$ for the momentum-dominated regime (S12 model) and constants $\pi_0$ and $c_3$ for the mass-dominated regime (S23 model). The questions are what similarity model approximates better the given flow and what is the best choice of the parameters. For S12 model, the balance of momentum

$$F + J = F_o, \qquad (80)$$

is applied to the domain with the inner radius $R_1 = D/2$ equal to the half diameter of the origin $D$ and the outer radius $R_2 = L$ equal to characteristic scale $L$ (see Fig. 7a). To approximate flow fields obtained by CFD simulation, which are shown in Fig. 6, ratio $L/D = 16$ is chosen. Figure 8 shows the left hand side of Eq. (80) in dimensionless form, which is a function of constant $c$. The value of $c$ corresponding to the given value of $F_o$ can be evaluated directly from Fig. 8. More accurate values listed in Appendix C are obtained by numerical solution of Eq. (80) with function $F$ calculated by interpolation using the data of Appendix B. Figure 5b shows that at $c > 15$ flux $F$ can be approximated as

$$\frac{F}{\rho v^2} = \frac{2450}{c_0 - c}. \tag{81}$$

This fit can be used in the momentum balance, Eq. (80), at $15 < c < c_0$.

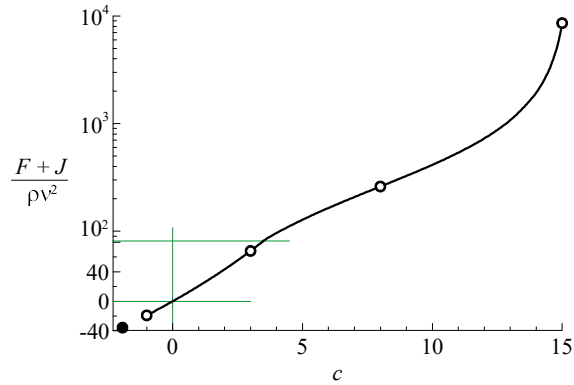

**Fig. 8.** Dimensionless total momentum flux $F + J$ for the similarity momentum-dominated jet as function of constant $c$. The closed point designates the solution, which does not belong to the curve. The open points on the curve indicate that there is no solution at these points

Appendix C lists the solutions of Eq. (80) at selected values of Reynolds number Re and normalized force $F_o$ obtained as the result of CFD simulation (see Table 1). The Reynolds number is essentially a normalized mass flux $M$. Note that momentum flux $J$ from the axis to the wall is greater than flux $F$ through the hemispherical surface at Re $\leq 20$ (see Appendix C). The relative contribution of flux $F$ increases with Re. This term dominates in the momentum balance at Re $\geq 50$. At Re $\geq 100$, flux $J$ is not greater than 3% of the total flux $F_o$, and it can be neglected in Eq. (80).

Parameters $\pi_0$ and $c_3$ of the similarity solution dominated by mass flux (S23 model) are related by mass balance, Eq. (68). The second equation can be momentum balance

$$J_a + J_w = F_o, \tag{82}$$

applied to the domain with the inner radius $R = D/2$ equal to the half diameter of the origin $D$ (see Fig. 7b). The solution of Eqs. (68) and (82) is

$$\frac{\pi_0}{v^2 D} = -\frac{1}{8}\left(\frac{1}{\pi}\frac{F_o}{\rho v^2} + \text{Re}\right). \tag{83}$$

$$\frac{c_3}{\nu D} = \frac{1}{32}\left(\frac{1}{\pi}\frac{F_o}{\rho \nu^2} - \text{Re}\right). \tag{84}$$

The calculated normalized values $\pi_0$ and $c_3$ along with momentum fluxes $J_a$ and $J_w$ are listed in Appendix C. Axial momentum source $J_a$ is lower than the flux through the wall $J_w$ at the tested values of Re ≤ 20. The relative contribution of $J_a$ into the momentum balance decreases with decreasing Re. This indicates that the term with singularity at the axis in Eq. (66) responsible for $J_a$ becomes less important if Re decreases.

Figure 6 compares flow fields calculated by CFD, S12, and S23 models in the range of Re between 2 and 200. At Re = 2, S23 approximates CFD at the parameters listed in Appendix C. However, a better agreement is found for S23 solution without singularity at the axis at $c_3 = 0$ (see the parameters in Table 1). Finally, this regular S23 solution is shown in Fig. 6 at Re = 2. At Re = 10, the CFD fields are satisfactorily approximated by the S23 solution with the parameters found from the momentum balance (Appendix C, the same parameters in Table 1). At Re ≥ 30, CFD fields can not be approximated by S23 model. However, a good agreement is observed with model S12 with the parameters evaluated from the momentum balance. The agreement is generally better at the near-wall region than around the axis. At Re ≥ 100, the CFD flow can deviate from model S12 at the distance from the origin greater than 10 (see, for example, contour lines $u_z = 0$ in Fig. 6d). This can be due to the influence of the external boundaries of the calculation domain. At Re ≥ 100, the CFD calculations show standing pulsations along the axis not described by S12 model. Contour line $p = 0$ (see Fig. 6b) can be chosen as the boundary separating the regions of the near-axis inner jet flow and the outer entrainment flow. The divergence angle of the inner jet flow decreases with Re.

Figure 6 shows the best agreement between the CFD and S12 models at Re = 50. The greater difference between the models at Re = 100 and 200 can be explained by errors at CFD calculation. The first error is mentioned above as the influence of the external boundaries of the calculation domain at $r > 10$ or $z > 10$. The second possible error is the scheme viscosity due to approximation errors, which becomes more important with increasing Re. The general trend is that model S12 becomes more precise with increasing Re. This is why one can suppose that model S12 evaluates the flow fields at Re > 200 too. On the contrary, model S23 becomes more precise with decreasing Re. According to the comparison between S23 and CFD, the regular S23 solution with $c_3 = 0$ can be applied at low Re ≤ 2.

Finally, the following algorithm can be proposed to find the best similarity solution approximating a jet emerging into a half space with the no-slip boundary condition. The input parameters are mass $M$ and momentum $F_o$ fluxes through the jet origin with diameter $D$ along with gas density $\rho$ and kinematic viscosity $\nu$. First, the jet Reynolds number is calculated according to Eq. (60) as

$$\mathrm{Re} = \frac{4M}{\pi \rho \nu D}. \tag{85}$$

If Re ≤ 10, S23 model should be used with parameters $\pi_0$ and $c_3$ calculated by Eqs. (83),(84). No satisfactory similarity solution can be generally found in the interval 10 < Re < 30. At Re ≥ 30, S12 model should be applied with parameter $c$ calculated from Eq. (80). The left hand side of this equation is shown as function of $c$ in Fig. 8. Essentially, this figure defines the dependence of $c$ versus $F_o$.

*4.4. Stresses at the surface*

Stresses at the wall are specific for the half-space problem in contrast to the Landau-Squire flow. Shear stresses are specific for the no-slip boundary condition. Previous studies of the half-space problem [29,30,32,33] devoted much attention to the stresses at the wall because they considerably contribute to the momentum balance. The same is confirmed by the present study. Section 4.2 reveals relations between the stresses and the parameters of S12 and S23 similarity models and calculates the associated momentum fluxes through the wall. Section 4.3 uses these fluxes to find the best parameters of the similarity solutions. Below, applied aspects of mechanical interaction between the fluid and the wall are considered. Studying stress distributions (see Fig. 9) is an additional way to compare and validate the models.

For S12 model, the normal and shear stresses at the wall are the same and given by Eq. (70). For S23 model, the normal stress is given by Eq. (77) and the shear stress is

$$\tau = -\frac{8\rho \nu c_3}{R^3}. \tag{86}$$

Note that $R = r$ at the wall. Figure 9 compares the stresses calculated by CFD with S12 and S23 models. At Re ≤ 10, the stresses decrease with the distance from the center as $1/r^3$ according to S23 model. Table 2 lists the parameters of S23 profiles shown in Fig. 9. Two different parameter sets are used at Re = 2. Model S23 agrees with CFD for the normal stress distribution (see Fig. 9a) and the shear stress at Re = 10 but disagrees with the shear stress at Re = 2 (see Fig. 9b). At Re ≥ 20, the stresses decrease as $1/r^2$ according to S12 model. The normalized profiles shown in Fig. 9 approach to the limiting curve corresponding to the maximum possible value of $c = c_0$ (dotted line) with increasing Re. In particular, this means that the stress distribution over the wall tends to a finite limit as momentum flux of the jet $F$ tends to infinity.

Table 2. Parameters of similarity model S23 matching CFD profiles of the stresses at the wall

| Case | Re = 2, $r < 4$ | Re = 2, $r > 4$ | Re = 10 |
|---|---|---|---|
| $-\pi_0/(\nu^2 D)$ | 0.654 | 0.5 | 5.47 |
| $c_3/(\nu D)$ | 0.0386 | 0 | 0.742 |

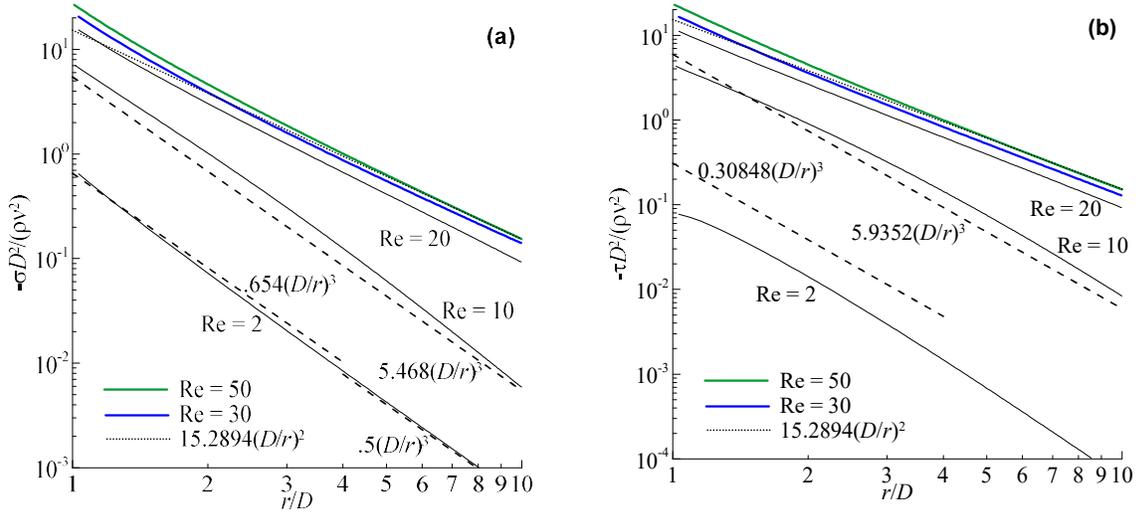

**Fig. 9.** Normal $\sigma$ (a) and shear $\tau$ (b) stresses at the wall: CFD (full lines); S23 model (dashed lines); S12 model at Re $\to \infty$ (dotted lines)

Figure 9 proposes that S12 model underestimates the CFD stresses at small $r$ at Re $\geq$ 30. The probable reason is neglecting the size of the jet origin (orifice or evaporation spot) in S12 model. Figure 6 shows that CFD levels are shifted relative S12 and S23 levels by the value around $D/2$. A shift $s$ can be introduced into Eq. (70) to improve the approximation of the stress profiles by S12 model,

$$\sigma = \tau = -c\frac{\rho v^2}{(R-s)^2}. \qquad (87)$$

Figure 10 shows the shear stress in the central zone around the jet origin. This zone is the most important because the stresses attain maxima there. CFD shear stress (solid lines) is well approximated by Eq. (87) (dashed lines) with the values of $s$ listed in Table 3. The values of $c$ are not changed. They are listed in Appendix C. Besides, Fig. 10 shows that Eq. (87) is applicable at $r/D >$ 0.6 at Re = 50. The shear stress distribution at Re = 50 is very close to the limiting distribution at Re $\to \infty$ (see Fig. 9b). This is why one can expect that the same domain of applicability and the same value of $s/D$ are valid for higher Reynolds numbers (see the right column in Table 3). The upper dashed line in Fig. 10 shows the estimated limiting profile of shear stress.

Table 3. Parameter $s$ of corrected S12 model to approximate CFD profiles of shear stress at the wall

| Re | 20 | 30 | 50 | $\infty$ |
|---|---|---|---|---|
| $s/D$ | 0.1 | 0.2 | 0.22 | 0.22 |

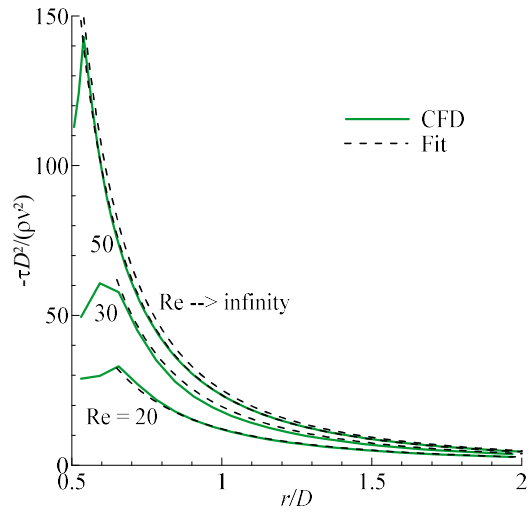

**Fig. 10.** Shear stress $\tau$ at the wall in the central zone around the jet origin: comparison between CFD calculations (CFD) and Eq. (87) (Fit)

*4.5. Application to laser evaporation*

Gas-phase jet flows are often observed at laser processing of materials. Laser beam can locally overheat the treated sample above the boiling point. Thus, a vapor jet is formed. Recent experimental studies of the powder-bed 3D printing technique by selective laser melting (SLM) [11,12,38-40] indicated that the entrainment flow of the vapor jet transfers powder particles to the melt pool and liquid droplets from the melt pool. Thus, the gas-phase flow substantially contributes to the mass transfer in the laser-interaction zone at SLM.

A model problem is solved for laser evaporation of iron target in argon ambient gas at atmospheric pressure $p_a$ = 1 atm. and room temperature $T_a$ = 298 K (normal conditions). It approximately corresponds to experiments on selective laser melting (SLM) [12]. Figure 11 shows a sketch of such an experiment. Thin laser beam scans a thick metallic substrate covered with a thin layer of powder with the thickness of 50-100 μm. The scanning laser beam forms a moving melt pool, which is partly evaporated. Behind the pool, a solidified bead of remelted powder is formed. The bead is surrounded with a thin zone free of powder, the so-called denuded zone. Tables 4 and 5 list the accepted laser beam parameters and material properties, respectively.

Table 4. Parameters of the laser beam [12]

| Wavelength | Diameter, $D$ | Power, $P$ |
|---|---|---|
| 1.064 μm | 100 μm | ≤ 170 W |

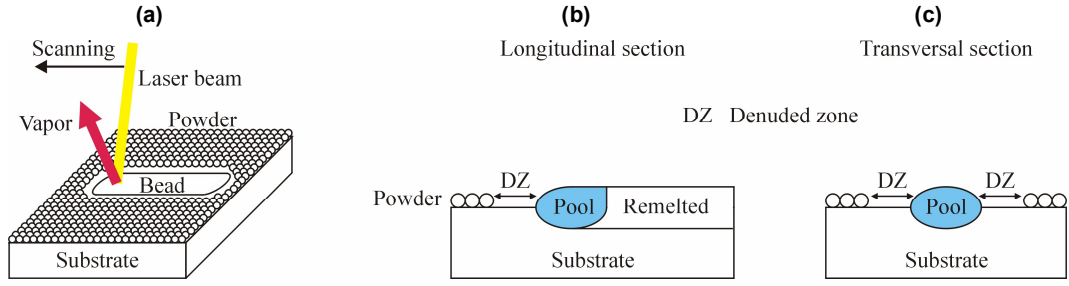

**Fig. 11.** Experiment on selective laser melting: general view of the sample (a) and longitudinal (b) and transversal (c) sections of the laser interaction zone

Table 5. Properties of materials for estimating the entrainment flow field at laser evaporation [41]

| Ambient gas Ar | Density at the normal conditions, $\rho$ | 1.784 kg/m$^3$ |
|---|---|---|
| | Kinematic viscosity at the normal conditions, $\nu$ | 12.7 mm$^2$/s |
| Target Fe | Atomic mass, $m$ | 56 a.m.u. |
| | Boiling point, $T_b$ | 3160 K |
| | Latent heat of evaporation, $Q$ | 6.09 MJ/kg |
| | Absorptance at 1.064 µm, $A$ | 0.3 |
| | Adiabatic exponent of vapor, $\gamma$ | 5/3 |

The incident laser energy is principally balanced by the losses by reflection, convective and conductive heat transfer in the target, and evaporation. Evaporation starts when the surface temperature attains the boiling point $T_b$. The pressure, temperature, and velocity of vapor are formed in the near-surface Knudsen layer and are related with the surface temperature and the saturated vapor pressure by the gas-dynamic conditions of strong evaporation [42]. Modeling of laser evaporation [12,43] indicates that the vapor velocity sharply increases with the laser power and attains the maximum possible value of the sound speed in the vapor. Such conditions of overheating with formation of vapor jet with the Mach number equal to one are typical for laser processing of materials [12,43]. Therefore, the vapor temperature and flow velocity are estimated, respectively, as the boiling point $T_b$ and the sound speed of Fe at the boiling point,

$$C = \sqrt{\gamma \frac{kT_b}{m}}. \qquad (88)$$

where $\gamma$ is the adiabatic exponent, $k$ the Boltzmann constant, and $m$ the atomic mass. The diameter of the evaporation spot is accepted to be equal to the beam diameter $D$.

According to the above assumptions, the estimated values of mass $M$, momentum $F$, and energy $P_v$ fluxes transferred by the vapor are

$$M = \frac{mp_a}{kT_b} C \frac{\pi D^2}{4}, \quad F = MC, \quad P_v = MQ, \qquad (89)$$

where $Q$ is the latent heat of evaporation. Table 6 summarizes the vapor parameters. At the maximum laser power $P = 170$ W, the evaluated energy loss by evaporation $P_v$ is a small fraction of the absorbed laser energy $AP = 51$ W, where $A$ is the absorptance taken from Table 5. Thus, the values of fluxes listed in Table 6 are more likely inferior bounds. The momentum flux is normalized by the values of density and viscosity of argon listed in Table 5 because the entrainment flow of argon gas is studied. Following the discussion in Section 4.3, at such a high value $F/(\rho v^2)$, the flow field can be approximated by S12 model and the contribution of flux $J$ into the energy balance, Eq. (80), is negligible. Therefore, the value of parameter $c$ listed in Table 7 is calculated by Eq. (81). The difference between $c$ and the maximum possible value $c_0$ is very small. Thus, the limit regime of entrainment flow is essentially attained. Further increase of momentum flux $F$ can not substantially change the flow field. In Table 10, ratio $c_2/c_1$ is calculated from $c$ by the method described in Section 2.2. The difference between this ratio and the corresponding value for the limit regime (see Table 1) is very small too.

Table 6. Estimated vapor parameters at the evaporation spot

| Flow velocity, $C$ | Mass flux, $M$ | Momentum flux, $F$ | Energy flux, $P_v$ |
|---|---|---|---|
| 884 m/s | 1.500 µg/s | 1.326 mN | 9.13 W |

Table 7. Parameters of S12 model for the laser evaporation

| $F/(\rho v^2)$ | $c$ | $c_2/c_1$ |
|---|---|---|
| $4.61 \cdot 10^6$ | 15.289 | -226.5168620 |

From the top view (Fig. 12a) one can see that powder particles on the substrate move radially toward the evaporation spot. Figure 12b shows a snapshot of the vapor jet. The camera was directed opposite to the scanning direction of the laser beam (see Fig. 11a), so that the experimental image corresponds to the scheme of Fig. 11c. The substrate is not visible in Fig. 12b. The top level of the substrate is approximately at the bottom of the 2 + 2 mm vertical scale bar. The vertical vapor jet is visible due to its thermal radiation. Elongated traces of particles visible at the top of the jet in Fig. 12b indicate the vertical direction of velocity. The velocity of these particles is measured as several meters per second [12]. The theoretically calculated velocity field (Fig. 12b) shows the same flow structure of the axial jet and the surrounding entrainment flow as the experimental images. Besides, the calculated tangent velocity near the substrate is around the experimentally measured velocity of particles on the surface [12].

The model S12 considerably underestimates the diameter of the inner jet flow. This is why it is useless to predict the axial velocity of the inner flow. The first reason of this disagreement is that S12 model neglects the size of the evaporation spot. The second reason can be probable turbulent character of the inner flow. This is consistent with experimentally observed considerable non-steady pulsations of the jet [12].

Figure 13a shows the top view of the laser-treated substrate. In the center, one can see the bead formed by remelted powder (compare with Fig. 11a) and the denuded zone around it. The dark domains on the left and right are the undisturbed powder layer. The axis of the bead is the line along which the laser beam scans the substrate. Figure 13b shows the profile of shear stress from the outer entrainment flow to the target surface, which is calculated by Eq. (87). The accepted value of shift $s = 22$ μm is estimated from Tables 3 and 4. The width of the denuded zone can be estimated theoretically by considering the balance of forces acting on a powder particle on the substrate surface (see Fig. 14). The drag force from the entrainment flow to the particle is approximately the product of shear stress $\tau$ and the projected area of the particle

$$F_D = \tau \pi d^2/4 , \qquad (90)$$

where $d$ is the particle diameter. The friction force opposed to the drag force is a fraction of the gravity force,

$$F_f = \mu \rho_p g \pi d^3/6 , \qquad (91)$$

where $\mu$ is the friction coefficient, $g$ is the gravitational acceleration and $\rho_p = 15$ g/cm³ the density of powder material (WC-Co).

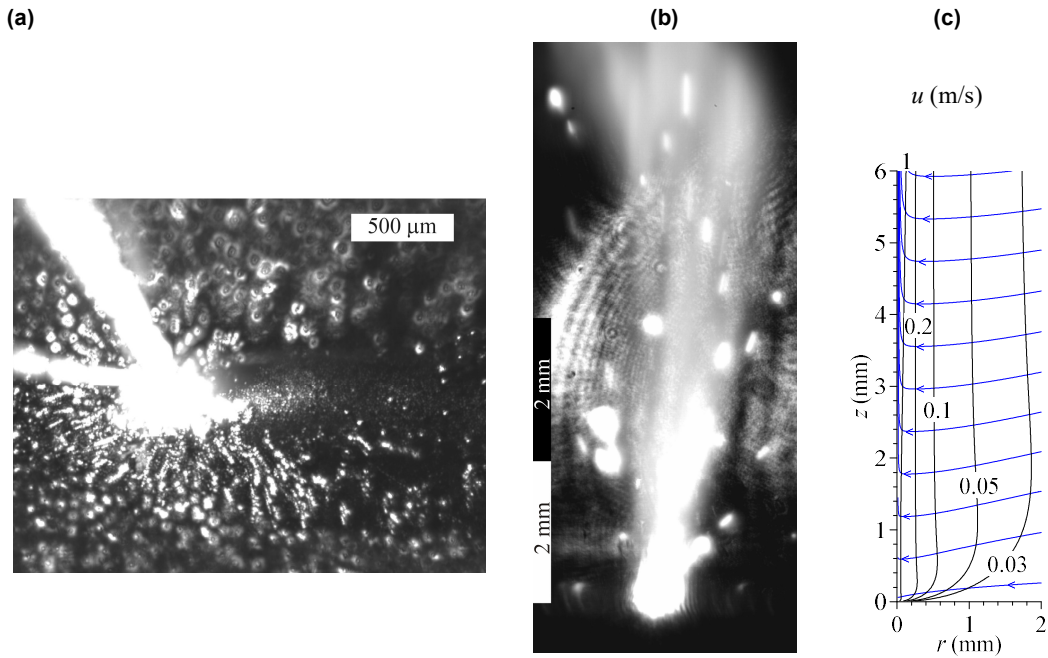

**Fig. 12.** Vapor jet at the laser experiment: (a) Particles on the substrate move toward the evaporation spot with the velocity up to 0.25 m/s [12]; (b) Vertical jet is formed; (c) Calculated by S12 model streamlines (curves with arrows) and contours of velocity absolute value $u$

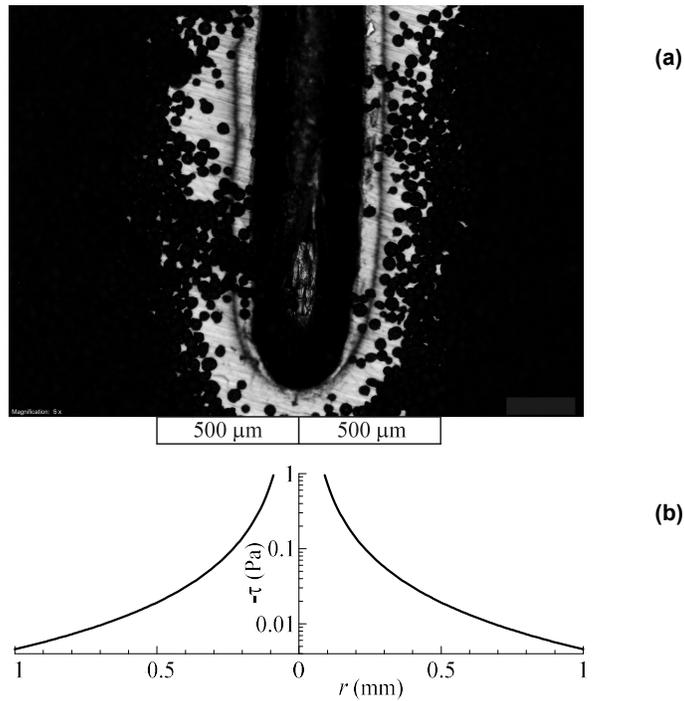

Fig. 13. Denuded zone around the laser-remelted bead (a) compared with the shear stress profile $\tau$ calculated by S12 model (b)

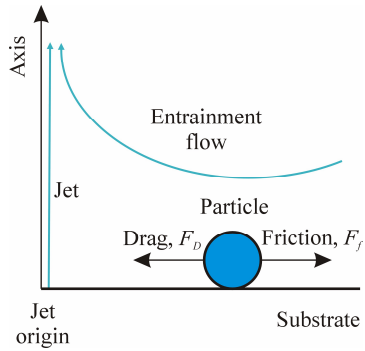

Fig. 14. Force balance for a powder particle on the substrate surface at SLM

Table 8 estimates the shear stress $\tau_0$ at the boundary of the denuded zone from balance $F_D = F_f$. Powder particles can undergo combined translational-rotational motion. This is why the value of friction coefficient $\mu$ (see Table 8) can be intermediate between the coefficients of sliding friction and rolling resistance [44]. Consequently, a wide range of $\tau_0$ is estimated (see Table 8). The distance $R$ at which shear stress $\tau$ is equal to $\tau_0$ estimates the half-widths of the denuded zone. It can be found graphically from the stress profile plotted in Fig. 13 or by solving Eq. (87),

$$R = s + \sqrt{\frac{c v \eta}{|\tau_0|}}, \qquad (92)$$

resulting $R = 150 .. 400$ μm. The experimentally observed half-width of the denuded zone (see Fig. 13) is around $300 .. 400$ μm. It is within the theoretically estimated range. Thus, the gas-dynamic mechanism for formation of the denuded zone at SLM seems to be reasonable.

Table 8. Powder parameters and the estimated shear stress $\tau_0$ at the boundary of the denuded zone

| Particle diameter, $d$ | 50 μm |
|---|---|
| Density (WC-Co), $\rho_p$ | 15 g/cm$^3$ |
| Friction coefficient, $\mu$ | 0.01 .. 0.1 |
| Boundary shear stress, $|\tau_0|$ | 0.05 .. 0.5 Pa |

Experiments on SLM at various pressures indicated that the denuded zone widens with decreasing pressure [39]. Equation (92) explains this result. It is known that dynamic viscosity $\eta$ of gases is approximately independent of pressure while kinematic viscosity $\nu$ is proportional to the mean free path, which is inversely proportional to pressure [45]. Thus, according to Eq. (92) increasing the kinematic viscosity with decreasing the pressure determines widening the denuded zone.

## 5. Conclusion

Two following similarity models of viscous flow are studied for emerging jet in a half-space with the no-slip boundary condition: S12 model results in exact solutions of the Navier-Stokes equations dominated by momentum flux while S23 model gives flow fields dominated by mass flux in the Stokes approximation. Physically meaningful solutions with the finite values of mass and momentum fluxes are found.

A method is developed to obtain S12 solutions for the half-space problem. The found S12 solutions have a singularity at the axis and can be presented as a set depending on parameter $c$ directly related with the distributions of normal and shear stresses over the wall. The physically meaningful solutions exist at the values of constant $c$ approximately equal to -1.93285 and in the range from -1 to $c_0 \approx 15.28938942$. Solutions are not found at the values of $c$ equal to -1, 3, 8, and 15. If a solution exists at the given value of $c$, it is unique.

Two linearly independent similarity solutions are found for S23 model. The first solution has singularity at the axis. The second one is regular in the half space.

Algorithms to evaluate the parameters of the similarity models are developed. The algorithms are based on the mass and momentum balances. The input parameters are the mass and momentum fluxes at the jet origin. Functional relations are obtained between parameter $c$ of S12 model and the momentum flux and two parameters of S23 model and the mass and momentum fluxes.

Comparison with CFD simulation for the viscous jet emerging from a circular origin (orifice or evaporation spot) indicates that S23 model is satisfactory at Reynolds numbers Re $\leq$ 10 and S12 model is satisfactory at Re $\geq$ 30. At Re $\leq$ 2, the regular S23-solution can be used. The parameter $c$ of S12 model tends to $c_0$ at increasing Re. In particular, this means that the stress distribution over the wall tends to a finite limit as the momentum flux transferred by the jet tends to infinity.

The obtained solutions for entrainment flow around a jet in a half-space explain formation of the denuded zone at selective laser melting. The theoretically estimated flow velocity corresponds to the experimentally observed velocity of particles transferred by the flow. The width of the denuded zone is estimated from the force balance. The estimated width corresponds to the experimentally observed one. Besides, the theory explains the experimentally observed widening of the denuded zone with decreasing the pressure of the ambient gas.

The methods proposed in this work can be generalized for conical domains with arbitrary half-angles and for other boundary conditions.

**Acknowledgement**

This work is supported by Russian Science Foundation (Grant Agreement No. 15-19-00254). The author is indebted to Roman Khmyrov for the experimental data shown in Figs. 12 and 13.34

**Appendix A.** Parameters $c$ and $c_2/c_1$ at which the no-slip and regular outside the axis S12 solution exists in a half space

| $c_2/c_1$ | $c$ | | | |
|---|---|---|---|---|
| | Branch I | Branch II | Branch III | Branch IV |
| -1000 | | 3.002622446 | 8.015898941 | 15.07169805 |
| -226.17066 | | | | 15.28938942 |
| -10 | | 3.232163640 | 8.821456134 | Points |
| -1 | | 3.801374483 | 9.274764959 | -1.932853446 |
| -.99 | -.9974657521 | | | |
| -.9 | -.8257586430 | | | |
| -.7 | -.3600988653 | | | |
| -.5 | 0 | 3.890372144 | 9.309303751 | |
| -.2 | .4165012170 | 3.949139647 | | |
| 0 | .6409573918 | 3.990526827 | 9.344955289 | 0 |
| 1 | 1.410898294 | 4.221478857 | | |
| 10 | 2.720619388 | 6.186874964 | 10.25742966 | |
| 1000 | 2.997372460 | 7.983899449 | 14.92457071 | |

**Appendix B.** Dimensionless momentum flux $F$ through the hemisphere around the origin for S12 solutions

| $c$ | $F/(\rho v^2)$ | $c$ | $F/(\rho v^2)$ |
|---|---|---|---|
| -1.93285 | 6.36270 | 5 | 18.8866 |
| -0.999990 | 2.67060 | 7.99984 | 85.7717 |
| 0 | 0 | 8.00016 | 85.7834 |
| 1 | -1.08843 | 12 | 465.721 |
| 2 | -0.264114 | 14 | 1629.11 |
| 2.99974 | 2.908259 | 14.999264 | 8236.65 |
| 3.00262 | 2.92133 | 14.9999264 | 8256.11 |

**Appendix C.** Parameters of similarity models S12 and S23 evaluated from the jet parameters

| Jet parameters | Re | 2 | 10 | 20 | 30 | 50 | 100 | 200 |
|---|---|---|---|---|---|---|---|---|
| | $F_o/(\rho v^2)$ | 10.16 | 106 | 396 | 882 | 2450 | 9900 | 39600 |
| S12 | $c$ | - | 4.33 | 9.82 | 12.53 | 14.26 | 15.04 | 15.23 |
| | $J/(\rho v^2)$ | - | 94 | 214 | 273 | 310 | 328 | 332 |
| | $F/(\rho v^2)$ | - | 12 | 182 | 609 | 2140 | 9572 | 39268 |
| S23 | $-\pi_0/(v^2 D)$ | 0.654 | 5.47 | 18.3 | - | - | - | - |
| | $c_3/(vD)$ | .0386 | 0.742 | 3.31 | - | - | - | - |
| | $J_a/(\rho v^2)$ | 1.94 | 37.3 | 167 | - | - | - | - |
| | $J_w/(\rho v^2)$ | 8.22 | 68.7 | 229 | - | - | - | - |